\documentstyle[12pt]{article}
\def\demi{{\textstyle {1\over2}}}

\def\L{l}

\def\pa{\partial}

\def\nn{\nonu}

\def\bea{\begin{eqnarray}}
\def\eea{\end{eqnarray}}

\def\nn{\nonu}

\def\be{\begin{equation}}
\def\ee{\end{equation}}
\def\s{\hat s}
\def\bg{\bar g}

\def\w{\wedge}
\def\o{\omega}
\def\t{\tilde}

\def\o{\omega}
\def\O{\Omega}
\def\L{ {\cal L}}

\let\nonu=\nonumber

\font\mybb=msbm10 at 12pt

\def\bb#1{\hbox{\mybb#1}}

\def\zet{{\bb{Z}}}

\def\real{{\bb{R}}}

\begin{document}
\bibliographystyle{perso}

\begin{titlepage}
\null \vskip -0.6cm
\hfill PAR--LPTHE 02--04



\vskip 1.4truecm
\begin{center}
\obeylines

         {\Large Topological Gravity versus Supergravity
                 on Manifolds with Special Holonomy}
\vskip 6mm
Laurent Baulieu$^a$, Alessandro Tanzini$^{a,b}$
{\em $^a$ Laboratoire de Physique Th\'eorique et Hautes Energies,
  Universit\'es Pierre et Marie Curie, Paris~VI
et Denis~Diderot,~Paris~VII}
{\em $^b$ Istituto Nazionale di Fisica Nucleare, Roma}

\end{center}

\vskip 13mm

\noindent{\bf Abstract}: We   construct  a topological theory for
  euclidean gravity in four
dimensions,
by enforcing self--duality conditions on the spin connection. The
corresponding topological
symmetry is associated to the
$SU(2)\times$diffeomor- \break phism$\times U(1)$
invariance. The
action of this theory is that  of
$d=4$, $N=2$ supergravity, up to a twist. The topological field theory
is $SU(2) \subset SO(4) $ invariant, but the full $SO(4)$ invariance
is recovered after untwist. This suggest that the topological gravity is
relevant for   manifolds with special holonomy.  The situation is
comparable to that of the topological Yang--Mills theory in eight dimensions,
for which the
$SO(8)$ invariance is broken down to $Spin(7)$, but is recovered
after untwisting the topological theory.
\vfill

\begin{center}

\hrule \medskip
\obeylines
Postal address: %
Laboratoire de Physique Th\'eorique et des Hautes Energies,
  Unit\'e Mixte de Recherche CNRS 7589,
  Universit\'e Pierre et Marie Curie, bo\^\i te postale 126.
4, place Jussieu, F--75252 PARIS Cedex 05

\end{center}
\end{titlepage}

\section{Introduction}

  Self-duality equations play an important role in the context of
topological quantum field theory (TQFT). They provide topological gauge
functions that one enforces in a BRST invariant way, which
determine in turn a supersymmetric action in a twisted form.
In~\cite{laroche}, there is a classification for possible
self-duality equations for the curvatures of forms of degree $p$ in
spaces with dimension $d\le 16$. For Euclidean gravity in four dimensions,
it has been known for a long time that the self-duality
equations solving the Einstein equations involve the
spin-connection itself~\cite{eguchi}. This apparent lack of
gauge independence of a physical equation is not harmful, since
its solution can be related to a physical solution - the
gravitational instanton- by a well-defined Lorentz
transformation.

In eight dimensions, an interesting result was found, namely
that the eight-dimensional topological Yang--Mills theory   gives a
$SO(8)$ covariant  theory which is
$Spin(7)$ invariant, but is a twisted version of the
eight-dimensional  supersymmetric Yang--Mills theory \cite{bakasi}
\cite{acharya} \footnote{By dimensional reduction to four dimensions and a
suitable gauge-fixing in the Cartan algebra allowed by topological
invariance, this theory gives  the abelian monopole theory of
Seiberg and Witten~\cite{swm}, which describes the infra-red
behavior of the non--abelian Super Yang--Mills theory
with extended supersymmetry \cite{bakasi}.}.
By analogy, we expect that topological gravities can  only be
relevant for manifolds with special holonomy. Indeed, the possible
gravitational self-duality equations for the spin connection
can only be invariant under
a subgroup of the Lorentz group (for instance
$SU(2) \subset SO(4)$ in four dimensions,
  $Spin(7) \subset SO(8)$ in eight dimensions). In the case of
topological Yang--Mills theories, there are arguments according
to which the full Lorentz invariance can be recovered, after
untwisting the topological field theory into a supersymmetric
theory. In the case of topological gravities, the situation is
more involved, because the gauge group is mainly the
Lorentz$\times$diffeomorphism symmetry which must be linked to local
supersymmetry instead of global supersymmetry. It is a priori not clear
whether the
untwisting procedure can be systematically  applied in a way that
transforms the topological gravity into its natural image,
$N=2$ supergravity.  This has been shown to be true in two
dimensions \cite{ba2gr}, and Anselmi and Fr\`e have given good
arguments that the four-dimensional $N=2$ supergravity can be twisted
in a topological gravity \cite{fre}.
There is moreover a quite general argument
that makes us confident that one can reach   all
$N=2$ supergravities, through the construction of topological
gravities coupled to TQFT's for $p$-form gauge fields.
Indeed, all types of superstring theories can be
obtained by suitable anomaly free untwisting of topological
sigma-models, which are quite easy to derive, at least in a formal
way~\cite{green}. Then,    we can   rely on the fact that
supergravities arise as low energy limits of superstrings in
order to predict a link between supergravities and topological
gravities.

The aim of this paper is to give a direct construction
of the topological gravity in four dimensions, and then
to show how it determines the
$N=2, d=4$ supergravity. As we will see, manifolds with special holonomy
play a relevant r$\hat{\rm o}$le both in the correct definition of the
topological theory and of the twist.
A nice feature of our work is that it
gives  a geometrical interpretation  of the graviphoton in
supergravity.
Another interest is that it
establishes a solid framework for generalizations in higher
dimensions. In the concluding remarks, we will comment
about a possible generalization of our topological action in eight
dimensions, which should be relevant for manifolds with $Spin(7)$
holonomy.

As for the determination of new invariants in four
dimensions, the TQFT is well defined on manifolds with an
$SU(2)$ holonomy group, as we will show in Sect.3.
These manifolds admit two covariantly constant chiral spinors.
The determination of observables stems from descent
equations, as it is standard in the BRST formalism, and describe a
cohomology which is equivariant with respect to the
$SU(2)\times$diffeomorphism symmetry. Their mean values can
only depend on global properties of the
chosen manifold for which one computes the action and the TQFT.

  \section{The topological gravity} Let us consider the symmetries
of ordinary gravity in four dimensions. The basic  symmetry is the
Lorentz$\times$diffeomorphism symmetry, whose gauge fields are the spin
connection
$\o^{ab}_\mu$ and the vielbein $e^a_\mu$. We consider these fields
as independent ones in a first order formalism, and through the
paper the latin indices $a,b,\ldots$ denote flat $SO(4)$ tangent
space indices, and $\mu,\nu,\ldots$ are world indices.  The
curvatures are
\bea
R^{ab}=d\o^{ab}+\o^{ac}\w \o^{cb} \ , \quad T^{a}=de^{a}+\o^{ac}\w e^{c} \ ,
\eea
where $D=d+[\o,\bullet]$ is the Lorentz covariant derivative. We define the Lie
derivative as $\L_\xi =i_\xi d- di_\xi$, where $\xi^\mu$ is an
anticommuting vector field. We could as well write the
following equation by redefining   $\L_\xi$ into the operator
$i_\xi D- D i_\xi$. The equations remain the same, up to a redefinition
of all ghosts and of ghost of ghosts, using the similarity operation~$\exp
i_\xi$ \cite{bb}.
  With the covariant
constraint
$T^a=0$, one has
$\o=\o(e)$, which determines
$\o$ up to a local Lorentz transformation. Eventually, a more refined
version of this constraint will be enforced in a BRST invariant way
in Section 5.

Since  the Lorentz$\times$diffeomorphism symmetry determines a Lie group,
the topological BRST symmetry can be straightforwardly constructed as:
\bea
se^a_\mu &=& \L_\xi e^a_\mu -\O^{ab} e^b_\mu +\Psi^a_\mu \nonumber \\
s\o^{ab}_\mu &=& \L_\xi \o^{ab}_\mu +D_\mu\O^{ab}   +\t \Psi^{ab}_\mu
\nonumber\\
\nonumber\\
s\Psi^a_\mu &=& \L_\xi  \Psi^a_\mu -\O^{ab} \Psi^b_\mu-\L_\Phi e^a_\mu
+\t \Phi^{ab}e^b_\mu
\nonumber\\
s \t \Psi ^{ab}_\mu &=& \L_\xi  \t\Psi ^{ab}_\mu -\O ^{ac}\t 
\Psi^{cb}  + D_\mu \t \Phi^{ab}
-\L _\Phi \o^{ab}_\mu
\nonumber\\
\nonumber\\
s  \Phi ^{a}  &=& \L_\xi   \Phi ^{a } -\O ^{ac} \Phi^{a}
\nonumber\\
s \t \Phi ^{ab}  &=& \L_\xi  \t\Phi ^{ab} -\O ^{ac}\t \Phi^{cb}
\nonumber\\
\nonumber\\
s\xi^\mu  &=& \Phi^\mu+\xi^\nu \partial _\nu \xi^\mu
\nonumber\\
s \O ^{ab} &=& \L_\xi  \O ^{ab}
-\O ^{ac}\O^{cb}  + \t \Phi^{ab} \ \ .
\label{brs}
\eea
Moreover, we have that $\Phi^a=e^a_\mu \Phi^\mu$, and thus, $s  \Phi
^{\mu} =
\L_\xi   \Phi ^{\mu } $.

Here the fields $\Psi$ and $\t \Psi$ are, respectively,  the topological
ghosts for 
$e$ and $\o$ and
$\Phi$ and $\t \Phi$ are their ghosts of ghosts.
We have that
\bea
s^2=0 \quad\quad   (s-\L_\xi )^2= \L_\Phi
\eea
on all fields.
These equations are the structure equations for topological gravity.
They can be cast in the following geometrical form, which generalizes
that given  in
\cite{symtop}:
\bea
\label{geo1}
&&(s+d) e^a +(\o^{ab}+\O^{ab}) e^b=\exp i_\xi (T^a +\Psi^a
+\Phi^a) \\
&&(s+d) (\o^{ab}+\O^{ab}) +(\o^{ac}+\O^{ac}) \w
(\o^{cb}+\O^{cb})=\exp i_\xi (R^{ab}+\t \Psi^{ab} +\t \Phi^{ab}).
\nonumber
\eea
It allows for the geometrical interpretation of all fields.

\section{ The graviphoton}
To determine a TQFT, we have to choose a topological gauge
function. The present paradigm is that the square of the gauge
function determines the bosonic part of the TQFT action, up to
boundary terms, as in \cite{laroche}. In the case of gravity in
four dimensions, there is a well known fact, which  one uses for
proving the positivity of the gravitational constant. It says
that, up to a pure derivative, the Einstein Lagrangian can be expressed
as a quadratic form in the Christoffel coefficients. In a first
order formalism, this translates into the following equation,  which
is satisfied by the
Einstein--Hilbert lagrangian:
\be\label{selfdual}
\L_{EH} = {1\over 4} \epsilon_ {abcd} \ R ^{ab} \w e^c \w e^d
  = \omega ^{ab^-}  \w  e^b \w  \omega ^{ac-} \w e^c
+d(\o^{ab^-}\w e^a \w e^b) \ .
\ee
For any given Lorentz tensor $X^{ab}$,  we define
   its  self-dual or antiself-dual
projection as
$X^{ab^\pm}=
\demi(X^{ab}\pm \demi \epsilon^{abcd} X^{cd})$.
Eq.~(\ref{selfdual}) is analogous to  the  equation $|F|^2= |F^\pm|^2
\pm F\w F$, which  allows one to express the Yang--Mills action as the
square of the self-dual part of the curvature, plus a boundary term.
However,  the  case of gravity is more subtle, since   the
decomposition in Eq.(\ref{selfdual}) is only
$SU_+(2)\subset Spin(4)$ invariant \footnote{Our notation
is $Spin(4)=SU(2)_+\times SU(2)_-$, where
$SU(2)_+$ has self-dual generators
$(\sigma^{ab})_\alpha^{~\beta}$ and $SU(2)_-$
anti-self-dual generators
$(\bar\sigma^{ab})_{\dot\alpha}^{~\dot\beta}$.}.
This is  related to the fact
that the  internal symmetry group cannot be disentangled from the
diffeomorphism symmetry, a situation that also complicates the
current ideas about the twist of supersymmetry.
Actually, in order that each term in the r.h.s. of Eq.~(\ref{selfdual})
is globally
well defined, one has to  define the vierbein
$e$ on manifolds with $SU(2)$  holonomy. Indeed, as we will see shortly,
this allows one to define transition
functions, such that $\omega ^{ab^-}  \w  e^b \w  \omega
^{ac-} \w e^c$, and  the   boundary term in Eq.~(\ref{selfdual}) can be
separately integrated to give a well defined action, which one can   
insert in the path integral.
We remark that the breaking of the full Lorentz symmetry into
a subgroup by the self-duality equation, and by the
decomposition of the action into a sum of a boundary term
plus a square,
is a phenomenon which also occurs  in the eight-dimensional
Yang--Mills theory, and in the more
general cases classified in~\cite{laroche}.

To define topological gravity, Eq.(\ref{selfdual}) suggests that
we choose the following topological gauge function - the gravitational
self-duality
equation - for the vielbein:
\bea
\label{ginstanton}
\omega _\mu ^{ab^-} (e)=0
\eea
  Any given solution of Eq.(\ref{ginstanton}) extremizes
the Einstein action, according to  Eq.(\ref{selfdual}). As explained in
\cite{eguchi}, although this equation is only $SU_+(2)$
invariant, it determines solutions of the $SO(4)$ invariant
equation $R^{ab^-}(\o(e))=0$.
Manifolds which admit metrics satisfying this condition have an $SU(2)_+$
holonomy group (they admit two covariantly constant chiral spinors);
then, the transition functions for the vierbeins $e^a_\mu$
can be chosen as elements of $SU(2)_+$. Correspondingly, by using the
torsion-free condition $T^a=0$ to define the $\o^{ab}=\o^{ab}(e)$, one can
easily show that the self-dual part $\o^{ab+}(e)$ transforms as a
connection, while the anti-self-dual part $\o^{ab-}(e)$ is globally
defined
under the $SU(2)_+$ local transformations. Thus each term on the r.h.s. of
Eq.(\ref{selfdual}) is well-defined on these manifolds.

We then conclude that the gravitational  TQFT
   that explores the fluctuations of the metrics around such
self-dual solutions  can  be meaningful only when it is defined over a
manifold with an $SU(2)$ holonomy group.  This is analogous to
what happens for the topological Yang--Mills theory in eight dimensions, for
which the manifold must have $Spin(7)$ or $SU(4)$ holonomy \cite{bakasi}.

The topological gravity will describe the moduli space associated
to Eq.(\ref{ginstanton}), which can be rewritten as :
\bea\label{ginstantons}
B^{ab^-} e^b \w  e^c \w \o^{ca} =0 \ \ ,
\eea
where $B^{ab^-}$ is a constant anti-self-dual Lorentz two-form.
$B^{ab^-}$  can be
constructed in terms of the two constant chiral spinors of the manifold.
Let us call
$\Psi=\delta e$  an element of the tangent bundle of the moduli
space of the gravitational instanton, solution of
Eq.(\ref{ginstantons}). By using the identity
$\delta (e^{[a}\o^{b]c} e^c) =d(e^{[a}\delta e ^{b]}) $,
one finds that $\Psi$ satisfies :
\bea
  B^{ab^-} d (e^{[a}\Psi^{b]})  = 0 \ \ .
\eea
Thus, in addition to the zero modes related to reparametrization
invariance,  we have the following zero modes for
$\Psi$ :
\bea\label{moduli}
  \Psi ^a \to \Psi^a + M^{ab^+} e^b,
\eea
where $M^{ab^+}(x) $ is an arbitrary self-dual local parameter.
The zero modes of Eq.(\ref{moduli}) reflect the $SU(2)_+$ invariance of
Eq.(\ref{ginstanton}), which determines the vierbein $e^a_{\mu}$ only up to
a local $SU(2)_+$ transformation.
Actually, Eq.(\ref{moduli}) has already  been captured in the BRST
symmetry equation for $\Psi$ in  Eq.(\ref{brs}), and it is the reminder
of this BRST equation after the breaking of the Lorentz symmetry down to
$SU_+(2)$ by  Eq.(\ref{ginstantons}), with $M^{ab^+}=\t\Phi^{ab^+}$.

One subtlety of topological gravity is the way we will gauge fix the
invariance in the ghost action corresponding to Eq.(\ref{moduli}).  Since
$M^{ab^+}$ is self-dual,  it corresponds to three degrees of freedom and
one must introduce Lagrange multipliers  that count for three independent
gauge-fixings. However, we can associate to
$M^{ab+}$
   a one-form defined modulo gauge transformation, which also counts
for~$3=4-1$~degrees of freedom. Thus, we will introduce an antighost
$A$ that is an abelian one-form, whose BRST variation is a fermionic
abelian one-form $\bar \Psi$.
Thus, the  very  basic
     invariance in   Eq.(\ref{moduli})
strongly suggest to introduce a
$U(1)$ graviphoton field  $A_\mu$ in topological gravity, with
the interpretation that $A_\mu$ has ghost number~$-2$ for consistency. The
action must only depend on
$A$ through $F=dA$, so that we expect that all ghosts associated to
the $SU(2)_+\times$diffeomorphism invariance carry no charge
under this additional $U(1)$ symmetry.

From a geometrical point of view, for  manifolds with a metric
satisfying Eq.(\ref{ginstanton}), the existence  of the zero modes
(\ref{moduli}) is related to the existence  of a certain number of
self-dual harmonic two-forms, equal to the Hirzebruch signature $\tau$ of
the manifold ${\cal M}$.
For the simplest example, the Eguchi--Hanson gravitational
instanton, one has only one  such self-dual two-form. In general,
one has that, locally, $M^{ab^+} e^a \wedge e^b = \sum_{i=1}^{\tau}
c_i~h^i$, where
$h^i$ is a basis of the cohomology group $H^{2+}({\cal M},\real)$.
Correspondingly, one has to introduce    a series of  Maxwell
fields $A^i$, $i=1,\ldots,\tau$, with ghost number~$-2$. This
   can be done by adding to the topological gravity action
  a sum of    topological Maxwell actions, which are constructed  in a
standard way from the  self-duality conditions on the  
curvatures $F^i=dA^i$ of the additional
$U(1)$ gauge fields.

The above considerations lead us to complete the BRST equations by
introducing the $U(1)$ graviphoton $A$. To do so, we must also
introduce an abelian ghost
$c$, a topological ghost $\bar\Psi$ and a ghost of ghost
  $\Phi$, so that:
\bea 
sA_\mu&=&    \L_\xi A_\mu+\pa_\mu  c +\bar \Psi_\mu
\nn\cr s\bar \Psi_\mu&=& \L_\xi \bar \Psi_\mu -\pa_\mu  \Phi
-\L_{\Phi}A_\mu
\nn\cr s \Phi &=&    \L_\xi  \Phi - \L_{\Phi} c
\nn\cr sc  &=&    \L_\xi c+ \Phi \ \ ,
\eea
and:
  \bea
\label{geo3}
(s+d) (A+c)  =\exp i_\xi
(F+\bar\Psi  + \Phi )
  \eea
The ghost numbers of the fields $ A,\bar\Psi, \Phi$, and $c$ are
$ -2,-1,0$ and $-1$, respectively.
Thus, the BRST symmetry that we
consider   corresponds to the topological
symmetry for the Lie algebra of the
$SU(2)_+\times$diffeomorphism$\times U(1)$ group. Since the
diffeomorphisms cannot be represented by finite matrices, we expect that
the untwist of the topological symmetry should determine
$N=2$ local supersymmetry, instead of global supersymmetry.

If one sets equal to zero
the ghosts $\xi^\mu$,  $\O ^{ab}$ and $c$ in all BRST equations, one
gets a BRST operator that is  nilpotent, modulo  a
diffeomorphism with parameter
$\Phi^\mu=e^\mu_a \Phi^a$, a Lorentz transformation with parameter
  $\t \Phi ^{ab}$  and a $U(1)$ transformation with parameter
$\Phi$. The cohomology of this operator define
the observables of the topological theory.

\section {The antighost sector}
In order to perform the topological gauge-fixing that realizes a
TQFT around the solution of Eq.(\ref{ginstanton}), we need to
complete the fields that have been introduced above, by introducing
suitable antighosts and lagrangian multipliers. Some of the latter will
actually be propagating fields, a situation that is a
current one in supergravity.
Moreover, only the $SU(2)_+$ sector of the Lorentz group, which is left
invariant by the gauge-fixing condition Eq.(\ref{ginstanton}), has to be
retained. The corresponding ghosts are given by the self-dual parts of
the fields
$\O^{ab}, {\tilde{\Phi}}^{ab}$ considered in Sect. 2.

To display the fields, we adopt
the presentation of  \cite{symtop}. In the following diagrams, the
ghost number is the same in any given column, and decreases by one
unit when one goes from one column to the next one on the right.
We have made the world indices explicit. Notice that $e$ and $\o$ have ghost
number zero, while
$A$ has ghost number -2; this implies in particular that the BRST doublet
$(\chi^{ab^+}, \beta^{ab^+})$ has positive ghost numbers equal to $(1,2)$
respectively.
  The following  triangular presentation  of
all fields can be justified if one introduces an additional
grading that accounts for the antighost number, which allows one
to generalize the geometrical equations  (\ref{geo1}) and
(\ref{geo3}). We have
\bea
\matrix
{  &    &    &    &  e^a_\mu  &   &   &   &
\cr
    &     &  \Psi^a_\mu  &   &    &  \bar\Psi^{ab^-}_\mu &    &
   &
\cr
    &   \Phi^a  &    &    &  L^{ab^-}, b^{ab^-}_\mu &
   &
\bar\Phi^a &
\cr
   &     &  \eta^{ab^-} &    &     &     \bar \eta^{a } &
    &
\cr
  }
\eea

\bea
\matrix
{  &    &    &    &  \o^{ab}_\mu  &   &   &
\cr
       &  &  \tilde\Psi^{ab}_\mu  &   &    & &
\bar{\tilde\Psi}^{ab}_\mu &    &
   &
\cr
    & {\tilde \Phi}^{ab^+} &    &    &  {\tilde L}^{ab^-},{\tilde
b}^{ab}_\mu &
   &   &
\bar{\tilde\Phi}^{ab^+} &
\cr
   &   &    {\tilde\eta}^{ab^-} &    &  &   &       \bar
{\tilde\eta}^{{ab}^+}
    &
\cr
  }
\eea

\bea
\matrix
{  &    &    &    &  A_\mu  &   &   & &  &
\cr
    &     &  \bar \Psi^{}_\mu  &   &   & &   {\chi}^{ab^+}
&    &
   &
\cr
    &    \Phi &    &    &  \beta^{ab^+}  &
   &   & &
  { \bar \Phi}  &
\cr
   &     &   &    &  &   &      { \eta}^{ } &
    &
\cr
  }
\eea

\bea
\matrix
{
           \xi^ \mu   &     &    \bar \xi^ \mu
\cr
     &    b^\mu
  }
\quad\quad\quad\quad\quad
\matrix
{
      \O^ {ab^+}  &   &        \bar \O^ {ab^+}
\cr
     &  b^{ab^+}  &
  }
\quad\quad
\quad \quad\quad\matrix
{     c   &   &        \bar c
\cr
    &  b   &
  }
\eea
Clearly, the antighost $\bar \Psi_\mu^{ab^-}$ and its Lagrange
multiplier $b^{ab^-}_\mu$ are introduced to implement the gauge condition
on $\o^{ab^-}_\mu$ as given by Eq.(\ref{ginstanton}), while ${\bar  {\t
\Psi}}_\mu^{ab}$ and its Lagrange
multiplier ${\t b}^{ab}_\mu$ will be used to impose, in BRST invariant
way, the relevant gauge condition on  the torsion $T^a_{\mu\nu}(e,\o)$,
which allows one to express the spin connection as a function of $e$,
and possibly of other fields. The rest of the fields are needed
because we are building a TQFT that is equivariant with respect
to the $SU(2)_+\times$diffeomorphism$\times U(1)$
symmetry. This means that
the gauge functions on the fields have an internal gauge symmetry, which
must fixed.  To do so, one needs suitable
ghosts for ghosts, antighosts
for antighosts and their Lagrange multipliers,
which allow one to remove the degeneracies of
all ghost propagators, such as the one  displayed   in
Eq.({\ref{moduli}).

All equations for the antighosts appearing in our field spectrum
are of the type
\bea
\s \bg = \lambda \quad  \s \lambda =\L_\Phi \bg + \delta_{\t \Phi} \bg
\eea
with  $s X= \s X +\L_\xi X+\delta_{{\O}^+} X$. Notice that
$\s^2 =\L_\Phi +\delta_{\t \Phi^+}$, where $\delta_{\t \Phi^+}$ is a
Lorentz transformation with parameter  $\t \Phi^{ab^+}$,
  and that none of the antighosts   transform under        the
$U(1)
$ symmetry  defined  in Eq.(\ref{geo3}).
We have :
\bea
\s\bar \Psi^{ab^-}_\mu =b_\mu ^{ab^-}
\quad  \s L^{ab^-}=\eta ^{ab^-}
\quad  \s \bar \Phi^{a }=\bar \eta ^{a }
\eea
\bea
\s\bar {\t \Psi}^{ab}_\mu ={\t b}_\mu ^{ab}
\quad  \s {\t L}^{ab^-}={\t\eta }^{ab^-}
\quad  \s \bar {\t\Phi}^{ab^+}=\bar {\t \eta} ^{ab^+}
\eea
\bea
\s  \chi^{ab^+} =\beta^{ab^+}
\quad  \s  \bar \Phi  =\eta
\eea
\bea
\s\bar \xi^{\mu} =b^{\mu } \ \ .
\eea
It is convenient to do field redefinitions
that express   the BRST transformation
for the antighost
$\bar\Psi^{{ab}^-}_\mu$ as follows:
\bea
\label{psielle}
\s \bar\Psi^{{ab}^-}_\mu &=& b_\mu ^{ab^-}
+ \pa_\mu L^{{ab}^-} \ \ , \\
\s  b_\mu ^{ab^-} &=&
(\L_\Phi + \delta_{\t \Phi^+})\bar\Psi^{{ab}^-}_\mu
+ \pa_\mu\eta^{{ab}^-} - [\t\Psi_\mu,L]^{{ab}^-} \ \ .
\nonumber
\eea
In this way one can take care in a more transparent way of the degeneracy
which occurs in the term
   $\bar\Psi^{{ab}^-}\w s(e^{[a}\o^{b]c}e^c)=
\bar\Psi^{{ab}^-}\w d(e^{[a}\w\Psi^{b]})$ of the action (\ref{ITQFT}).
The corresponding zero-modes  for $\bar\Psi^{{ab}^-}_\mu$ are explicit
  in Eq.(\ref{psielle}).

\section{The topological gravity action}

In order to determine  a TQFT for topological gravity,  which uses
the above fields, we need a little bit of thinking.  The expression
Eq.(\ref{selfdual}) of the Einstein action    and the
the self-duality equation (\ref{ginstanton}) are the signal for the
existence  of a TQFT for gravity. It is
quite clear that the definition of a TQFT that concentrates   in a
BRST invariant way the path
integral around the solution of    Eq.(\ref{ginstanton})  breaks the
$SO(4)$ invariance down to $SU_+(2)$. We must therefore consider a
quantum field theory in which the path integral measure only involves
vierbeins that  determine a globally well defined topological Lagrangian.
As discussed in Sect.3, this restricts the choice to manifolds
with $SU(2)$ holonomy.

  To recover the
full Lorentz symmetry after an untwisting procedure, the existence of the
topological ghost $\bar
\Psi$   of the $U(1)$ gauge field
$A$ is essential. It allows a complete  determination of  spinor
fields. Indeed, the fields
\bea\label{gravitino}
(\Psi^a_\mu, \bar  \Psi_\mu ^{ab^-}, \bar \Psi _\mu)
\eea
can be untwisted to
the two gravitini $(\lambda ^{\dot \alpha i}_\mu,
\lambda ^{  i}_{\mu \alpha} )$, where $i=1,2$ labels the
automorphism group~$SU(2)_R$ of $N=2$ supergravity. As
shown explicitely in the next section, this follows from
the   fact that,
two  pairs of dotted and undotted spinors can be twisted into a
vector, a selfdual or antiselfdual two-form and a scalar, thanks to
the two different possibilities of extracting an $SO(4)$ symmetry
from   the $SU_+(2)\times SU_-(2) \times SU(2)_R $ symmetry
group \cite{twist}.
Thus,
we need the Lorentz scalar ghost  $\bar \Psi_\mu $, and we  understand  
that
$A_\mu$ must have ghost number
$-2$, in order that $\bar \Psi_\mu$ has ghost number $-1$, as
$\bar \Psi^{ab^-}_\mu$. This completes the analysis done in the previous
section for the introduction of $A$.

In supergravity, the standard gauge condition for
the local supersymmetry on the spin 3/2 gravitini
$\lambda_\mu^i$ is
\be
\gamma^\nu D_\nu\gamma^\mu
\lambda_\mu^i=0 \ \ .
\label{panini}
\ee
If this condition is imposed in a way that respects  the ordinary   BRST
symmetry of supergravity, it  yields a propagation of the Lagrange
multipliers, as first observed by Nielsen and Kallosh
\cite{nk}.
This leaves little room,
but for the interpretation of the fields
\bea \label{gh}
(\Phi^a , L ^{ab^-}+    {\t L}^{ab^-} , \Phi )
\ , \
(\bar \Phi^a , L ^{ab^-}-    {\t L}^{ab^-} , \bar \Phi )
\eea
as the twist of   pairs of commuting Majorana spinors that will
be identified, eventually, as the  ghosts and antighosts of the local
supersymmetry of $N=2$ supergravity. Then,
\bea\label{lm}
(\bar \eta ^a , \eta^{ab^-}, \eta )
\eea
can be identified as the twist of anticommuting
Majorana spinors, which are the propagating Lagrange
multipliers for the gauge--fixing on the gravitini, as we will see in
more detail in the next section.
The twisted version of the gauge condition (\ref{panini})
naturally arises as the relevant gauge fixing for the topological ghosts
that occur in topological gravity.

This parallel is just enough to tell us how to choose the gauge
functions for having a theory which, (i), concentrates the path
integral around the gravitational instanton, and (ii),  has a BRST
symmetry corresponding to a twisted supersymmetry. Since the Lie algebra we start
from contains the reparametrization symmetry, which cannot be represented
by finite matrices,  we   expect  a link between
this topological  BRST symmetry and local     supersymmetry, rather than
  global supersymmetry.

We face the problem of determining an action that contains the
Einstein--Hilbert term, expressed  under the form displayed in
Eq.(\ref{selfdual}),  plus a term that
depends on the sixteen fermionic degrees of freedom contained in
Eq.(\ref{gravitino}). This action is :
\bea\label{ITQFT}
I_{TQFT}&=&
\int  s\Big[
{\bar \Psi}^{ac^-}\w  e^c\w e^b \w (\o^{ba^-} +b^{ba^-})\Big] \\
&&+ s\Big[\chi^{+}\Big(\beta^{+}+
2(F + \bar\Psi^{ab^-} \w\bar\Psi^{ab^-} +\bar\Psi\w\bar\Psi)
\Big) + 2 F^-\w\Psi^a\w e^a \Big] +
\nn\cr
&&+s\Big[\bar {\t \Psi}^{ab} \w e^b  \w^* (T^a(\o,e)
+\bar\Psi^{ab^-} \w \Psi^b +\bar\Psi\w\Psi^a)\Big] \ \ , \nonumber
\eea
where $\chi^+=\chi^{ab^+}e^a\w e^b$ and
$\beta^+=\beta^{ab^+}e^a\w e^b$. One must expand the $s$-exact term,
using the above definition of $s$, to get the full expression of
$I_{TQFT}$. Notice that the dependence on
   the Lagrange multiplier $\beta^{+}$,
which has ghost number 
two, arising from the second line of Eq.(\ref{ITQFT}) breaks the
$U(1)$ ghost number symmetry of this action down to a $\zet_2$ symmetry.
This means that the ghost number is conserved only modulo two.
This is an unavoidable feature if we want
the graviphoton field $A$ to propagate with a Maxwell term
$F\w^* F$, since, according to our previous definitions, this
term has ghost number $-4$.
 From the twisted supergravity point of view, this can be understood
as a consequence of the fact that the $U(1)_R$ group, which
  is identified with the ghost number symmetry after the twist, is only a
symmetry of the equation of motions of $N=2$ supergravity, and
not of the whole action \cite{fre}.
However, this problem can be avoided in the topological theory, which
can be alternatively defined  without the   term in
$(\beta^{+})^2$ responsible of the $U(1)$ symmetry breaking.
In this case, the $A$ field does not propagate, and the topological action
(\ref{ITQFT}) simply localizes this field to its classical solutions.

The last term in the Eq.(\ref{ITQFT})
gives a constraint that
determines  in a Lorentz covariant way the spin
connection
$\o$ as a function of $e$ and of other fields in Eq.(\ref{gravitino}).
It also allows one to eliminate the Lorentz topological ghosts $\t
\Psi$ and $\bar {\t \Psi}$ in function of the other ghosts
by algebraic equations of motion.


In order to fix the gauge for the   topological ghosts, we
consider the following $s-$exact term :
\be\label{gfghost}
I_{ghosts}=
\int d^4x \
s\Big[\sqrt{g}(\bar \Phi ^a   D_\mu   \Psi ^{a }_\mu+
{\t L}^{ab^-}  D_\mu   \bar \Psi ^{ab^-}_\mu
+\bar \Phi \pa_\mu   \bar \Psi_\mu)\Big]
\ee
This term, after untwisting, provides the propagation of
the fields in Eqs.(\ref{gh}) and (\ref{lm}), together with a
gauge-fixing for the longitudinal parts of all topological ghosts
in Eq.(\ref{gravitino}).

In order to establish the  comparison with the twisted
supergravity action, we need an  identification
of $\chi ^{ab^+}$ with
$\O ^{ab^+}$, which is obtained by the additional term:
\be
\label{id}
I_{\chi/\O}=
\int d^4x \
s\Big[{\sqrt {g} }\Big(\bar {\t \Phi}^{ab^+}
(\O^{ab^+} -\chi^{ab^+})\Big)\Big]  \ \ .
\ee
  In the next section we will see in fact that the field $\chi^{{ab}^+}$
does not appear in the twisted supergravity
multiplet. The term (\ref{id})  also gauge-fixes the symmetry in
Eq.(\ref{moduli}), by providing an equation for the ghost-for-ghost
$\t\Phi^{ab^+}$.  Notice that the terms in $\O^{ab^+}$ appearing in
Eq.(\ref{ITQFT}) after the identification (\ref{id}) can be
easily reabsorbed thanks to the $SU(2)_+$ equivariance
of the action\footnote{They can actually be reabsorbed
by the field redefinition
$\Psi^{a\prime}=\Psi^a+\demi \O^{ab^+} e^b$, which does not induce
additional terms in the action due to its $SU(2)_+$ equivariance.}.
In the next section we will show that the resulting
topological action corresponds
to the twisted version of $N=2$ pure supergravity.

As for the gauge fixing of the local Lorentz symmetry, we proceed in
two steps.
We know that, if the Riemann curvature
is self-dual, {\it i.e.} $R^{ab^-}=0$,   one can always find an
$SU(2)_-$ Lorentz transformation such that also the corresponding spin
connection is self-dual, $\o^{ab^-}=0$ \cite{eguchi}. The first line
of Eq.(\ref{ITQFT}) exactly implements this condition in an $SU(2)_+$
equivariant way. Eventually, the left-over $SU(2)_+$ invariance can   be
fixed by the term:
\be
I_{Lorentz}=
\int d^4x  \
s\Big[{\sqrt {g} }\Big(\bar {  \O}^{ab^+ }
(e^a_\mu \delta ^\mu _b -e_a^\mu \delta _\mu ^b)\Big)\Big] \ \ .
\ee

The   action
\be\label{total}
I= I_{TQFT}+I_{ghosts}+I_{\chi/\O}+I_{Lorentz}
\ee
is thus our  candidate for describing   topological gravity.
It is still $U(1)$ and reparametrization invariant.  The gauge fixing
of these last symmetries can be completed by the obvious term
$ s(\bar c \pa_\mu A_\mu +\bar \xi^\nu \pa_\mu g_{\mu\nu})$. A more
refined analysis of this gauge fixing term could be done, by introducing
the extended formalism with another BRST operation, which would  control
the reparametrization and $U(1)$ invariances, in a way that is analogous
to that used for other types of topological  gauge theories in \cite{lbz}.

As for the observables, the situation is as follows. We have constructed a
Lagrangian that is globally well-defined over any given four-dimensional
manifold
  ${\cal M}$  with $SU(2)$ holonomy. It can be considered as the
gauge-fixing of   topological invariants, which  can be  combinations
of
$Tr (R^{ab}\w R^{ab})$,     $Tr (\epsilon_{abcd} R^{ab}\w~ R^{cd} )$
and  $F\w F$. The operation $s$
describes a topological symmetry that is equivariant with respect to
$SU(2)_+$ and reparametrization symmetry.  Eq.(\ref{geo1})   indicates
that possible observables are the field polynomials obtained by doing
the substitutions
$T\to T +\Psi+\Phi$,    $R\to R+\t \Psi+\t \Phi$ and
$F\to F +\bar \Psi+\Phi$,
  in all invariant
polynomials in $R$, $T$ and $F$ that one can construct, as a
generalization of
\cite{bakasi}. The claim is that the expectation values of these
observables, computed from the action in Eq.(\ref{total}), only depend on
the differential structure of ${\cal M}$. The amazing feature is the
relationship of the whole action to Poincar\'e supergravity.

\section{The relationship to the supergravity action}

We will now see that the  topological gravity theory discussed in the
previous section corresponds to the twisted version of four dimensional
$N=2$ pure  supergravity. The twist of this theory has been already
considered in \cite{fre}. What we want to underline here is that 
special holonomy manifolds are required in order to have a well-defined
twist operation. We  have seen in the previous section that the
topological gravity lagrangian  is globally  well-defined on manifolds
with
$SU(2)$ holonomy. The  corresponding BRST algebra is associated with the
$SU(2)_+\times$diffeomorphism$\times U(1)$ symmetry, while the
remaining $SU(2)_-$ factor of the $Spin(4)$ group is a global symmetry.
Thus, it is possible to define the twist on such manifolds, as one does
in the  standard global supersymmetry case, by redefining a new
   factor $SU(2)_-^{\prime}$, such that:
\be
SU(2)_-^{\prime} \equiv {\rm diag} (SU(2)_-\oplus SU(2)_R) \ \ .
\ee
  Correspondingly, the fields of the topological gravity,
given by a vector, an
anti-self-dual two-form and a scalar
denoted as $(X^a, \bar X^{{ab}^-}, \bar X)$, determine two  dotted and
undotted spinors  according to :
\bea
X^{\alpha \dot\beta}&=&X^a (\sigma ^a)^{\alpha \dot\beta}
\nn \\
X^{~\dot\alpha}_{\dot\beta}&=&\bar X^{ab}
(\bar\sigma ^{ab})^{~\dot\alpha}_{\dot\beta}
+\bar X  \delta^{~\dot\alpha}_{\dot\beta}
\label{twist}
\eea
One assembles $X^{\alpha \dot\beta}$ and
$X^{~\dot\alpha}_{\dot\beta}$ as a pair of two Majorana spinors $\lambda
^i$,
by identifying the index $\dot\alpha$ in Eq.(\ref{twist}) as an
internal $SU(2)_R$ index $i$.
The dictionary between the fields of
topological gravity and of supergravity   is thus:
\bea
\label{twist1}
(\Psi^a_\mu, \bar  \Psi ^{ab^-}_\mu, \bar \Psi _\mu)
&\to&  \lambda ^i_\mu
\nn\cr
  (\bar
\eta ^a ,
\eta^{ab^-}, \eta ) &\to& d ^i
\nn\cr
(\Phi^a , L ^{ab^-}+    {\t L}^{ab^-} , \Phi )&\to&  \kappa^i
\nn\\\
(\bar \Phi^a , L ^{ab^-}-    {\t L}^{ab^-} , \bar \Phi )&\to&
\bar
\kappa^i
\eea
$\lambda ^i_\mu$ stand for two spin 3/2 gravitini, $\kappa^i$ and
  $\bar \kappa^i$ stand for two pairs of commuting Majorana
spinors that are the spin 1/2 ghosts and antighosts for
$N=2$ local supersymmetry, and $d^i$ are  the corresponding spin
1/2 anticommuting Lagrange multipliers.

If we restrict to terms that are not more than quadratic in the
$\Psi$ and $\bar \Psi$ topological ghosts, it is easy to get
that, after computing the $s$-exact terms in Eq.(\ref{ITQFT}), and
eliminating the fields with algebraic equations of motion,
\bea\label{quad}
I_{TQFT}
&\cong& \int \omega^{ab^-}\w e^b \w
\omega^{ac^-}\w e^c
+ \nn \\
&&-(d\Psi^a + \o^{ba^+}\Psi^b)\w\bar\Psi\w e^a + \nn \\
&&-(d\Psi^a + \o^{ba^+}\Psi^b)\w\bar\Psi^{ac^-}\w e^c + \nn\\
&&- F^+\w F^+ -
2 F^+\w (\bar\Psi^{ab^-}\w\bar \Psi^{ab^-} +\bar\Psi\w\bar\Psi)
+\nn\\
&&+ 2 F^- \w\Psi^a\w \Psi^a  \ \ .
\eea
Notice that in this action the covariant derivatives with respect
to the $SU(2)_+$ factor of the Lorentz group appear,
as   expected in  a construction that is  $SU(2)_+$ equivariant. Up to
the boundary term
$d(\o^{ab^-}\w e^a\w e^b)$ coming from the decomposition
(\ref{selfdual}), the action in
Eq.(\ref{quad}) yields, after untwisting, 
  the $N=2$ supergravity action, which
reads :
\bea\label{sugra1}
I_{SUGRA}
&\cong&\int {1\over 4} \epsilon_{abcd} R ^{ab} \w e^c \w e^d
+ D\bar\lambda ^i \w \gamma_5 \gamma^a \lambda^i \w e^a
- F^+\w F^+ + \nn\\
&&-{1\over 2} F\w\bar\lambda^i\w\gamma_5\lambda^j \epsilon_{ij}
-{1\over 4}
\epsilon_{abcd}F^{ab}e^c\w e^d\bar\lambda^i\w\lambda^j\epsilon_{ij}
  \ \ ,
\eea
up to quartic terms in the gravitini.
As for these terms, they should be taken into account
by a suitable redefinition of the connection $\o^{ab}$.
We recall in fact that in supergravity
there are extra contributions to the torsion, such that the
defining equation for the connection
is actually
\bea
&&de^a +\omega ^{ac}(e,\Psi,\bar \Psi) \w e^c+
\demi \bar \lambda ^i \gamma^a \w \lambda ^i=\nn\\
&&=T^a(\o,e)
+\bar\Psi^{ab^-} \w\Psi^{b} +\bar\Psi\w\Psi^a=0 \ \ .
\label{to}
\eea
Thus, the solution of Eq.(\ref{quad}) is a function
$\omega ^{ac}(e,\Psi,\bar \Psi)$, which  contributes to the presence in
the action of quartic terms in the fermionic fields. Such
quartic terms come from the Einstein action, as well as from the second
line in the topological action Eq.(\ref{ITQFT}).

This comparison allows us to truly identify  the
topological ghosts of $e$ and $A$ as fields that can be twisted
into two gravitini. Moreover the $U(1)$ gauge field  is the
graviphoton field. As already remarked in the  previous section, a subtle
feature is that $A$ has ghost number equal to minus two, which implies
that the $U(1)$ ghost number symmetry is broken
to $\zet_2$.


As for the rest of the action, the terms that depend on $L$  
  in Eq.(\ref{ITQFT}) and the Lagrangian   Eq.(\ref{gfghost})  provide
an action, which is,  after using the correspondence
Eq.(\ref{twist1}) :
\bea
\label{sugragf}
\int d^4x \sqrt{g}
(
d^i
\gamma^\nu
D_\nu
\gamma^\mu
\lambda ^i_\mu
+
\bar
\kappa^i
\gamma^\nu
D_\nu
\gamma^\mu D_\mu
\kappa ^i) \ \ .
\eea
This is precisely what is needed to fix local supersymmetry in
supergravity, and completely identifies
the ghost of ghosts,   antighosts of antighosts and their Lagrange
multipliers in Eq.(\ref{twist1}) as the twisted ghosts,
antighosts and Nielsen-Kallosh ghosts of supergravity.

\section{Remarks and conclusions}

It is striking that a straightforward construction of
topological gravity for manifolds with $SU(2)$ holonomy can be obtained as
the BRST invariant gauge-fixing of the standard invariants,
$R^{ab}\w R^{ab}$ or
$\epsilon_{abcd} R^{ab}\w R^{cd}$ and $F \w F$. We can
define observables by considering the cocycles that arise
from these invariants, owing to the
geometrical  equations (\ref{geo1}) and (\ref{geo3}). This possibility
was already proposed by Witten \cite{wittengravity} in the context of
Weyl gravity, and by Anselmi and Fr\`e \cite{fre}.

As for the utility of topological gravity, it seems that it
holds only for manifolds of a certain type, namely for those that
admit gravitational instantons, as hyper-K\"ahler manifolds.
As we have seen in Section 3, the fields of the topological gravity
multiplet have a nice geometrical interpretation in terms of linear
deformations around the gravitational instanton\footnote{An explicit
evaluation of the zero-modes of the Eguchi-Hanson gravitational instanton
can be found in
\cite{bfmr}.},  which seems quite analogous to that of the fields of the
topological Yang--Mills theory in  four dimensions for the gauge
instantons \cite{twist, bs}. The manifolds involved in our analysis admit
self-dual abelian connections, which can be naturally included in our
topological model.
We underline that the presence of such $U(1)$ connections for these
manifolds, also called "abelian instantons", makes it difficult to 
compute for these manifolds
the Seiberg--Witten topological invariants
associated to the twist of theories with $N=2$ global
supersymmetry~\cite{swm}. The topological gravity action, which  we have
discussed, could be  useful to get information on these cases.

It is however unclear to us whether one can give to the
four-dimensional boundary term  $d(\o^{ab^-}\w e^a\w e^b)$
any interpretation as a topological term.
This is maybe possible
on certain manifolds, {\it e.g.}, hyper-K\"ahler ones.
Were it the case, it would provide cocycles from
descent equations, which could be used tentatively to
compute new  invariants from the path integral.

A progress introduced by the present work is a rather systematic way
for having topological
gravities in higher dimensions.  A promising case is in 
eight dimensions, for which \cite{bakasi} suggests that one
can
use the octonionic gravitational instanton equation \cite{floratos}
as a generalization  of Eq.(\ref{ginstanton}),
\bea\label{octo}
\o^{ab}-\demi \O^{abcd} \o^{cd}=0
\eea
where $\Omega_4$ is the octonionic four-form used in \cite{bakasi}.

It is indeed well known that, on an eight-dimensional manifold,
one has a $Spin(7)$
invariant decomposition of two-forms, as ${\bf 28=7} \oplus {\bf 21}$,
corresponding respectively to the eigenvalues $\lambda = -3,1$ of the
octonionic four-form $\O_4$. Eq.(\ref{octo}) corresponds to
setting to zero the components of the curvature in the former seven
dimensional subspace. Following the same line of reasoning as in Sect.3,
one foresees the following  generalization of the four dimensional
lagrangian:
\bea
  \int_ {{\cal M}_4}\o^{ab^-} \w\o^{ac^-}\w e^b\w e^c
\to
\int_{{\cal}_{{\cal}_ {{\cal M}_8}} } \Omega_4\w \o^{ab^-} \w\o^{ac^-}\w
e^b\w e^c
\eea
This is a well-defined lagrangian on manifolds with $Spin(7)$ holonomy. It
is thus tempting to consider a cohomological theory, whose BRST symmetry
is associated to a $Spin(7)\times$diffeomorphism$\times$gauge invariance,
corresponding to the following
generalization of the four dimensional case
\bea
\label{eight}
  \int_{{\cal}_ {{\cal M}_4}} s (\bar \Psi ^{ab^-} \w(\o^{ac^-}+b^{ac^-})
\w e^b\w e^c)
\cr
\to
\int_{{\cal}_ {{\cal M}_8}} \Omega_4\w s (\bar \Psi ^{ab^-}
\w(\o^{ac^-}+b^{ac^-}) \w e^b\w e^c) \ \ .
\eea
Manifolds with $Spin(7)$ and $G_2$ holonomy have recently attracted  a
renewed interest (see, {\it e.g.}, \cite{sp7}).
The relevant r$\hat{\rm o}$le, which is
played in this context by the self-duality
conditions on the spin connection, has been
underlined in \cite{bilal}.
We are currently studying the link between the action (\ref{eight}) and that
of supergravity in eight dimensions. Moreover, the dimensional reduction of
the theory stemming from the action (\ref{eight})   provides very
interesting models in lower dimensions, in the spirit of \cite{bakasi}.
For example, the reduction to seven dimensions should be related to $G_2$
holonomy manifolds.
We can also expect an interesting low-energy effective theory for
$N=2$ supergravity in four, two and zero dimensions.

  {\bf Acknowledgments}:
We thank M. Bellon, I. Singer and R. Stora for very useful
discussions.


\end{document}